\documentstyle[aps,prl,twocolumn]{revtex}
\begin{document}
\draft
\title{Scaling Behavior in Turbulence is Doubly Anomalous}
\author {Victor L'vov, Evgenii Podivilov$^*$ and Itamar Procaccia }
\address{Department of Chemical Physics,
The Weizmann Institute of Science, Rehovot 76100, Israel \\
$^*$ Institute of Automation and Electrometry, Ac. Sci. of Russia,
630090,
Novosibirsk, Russia}
\maketitle
\begin{abstract}
It is shown that the description of anomalous scaling in turbulent
systems requires the simultaneous use of two normalization scales.
This phenomenon stems from the existence of two independent (infinite)
sets of anomalous scaling exponents that appear in leading order, one
set due to infrared anomalies, and the other due to ultraviolet
anomalies. To expose this clearly we introduce here a set of local
fields whose correlation functions depend simultaneously on the the
two sets of exponents. Thus the Kolmogorov picture of "inertial range"
scaling is shown to fail because of anomalies that are sensitive to
the {\em two ends} of this range.
\end{abstract}

Anomalous multi-scaling in turbulence is usually discussed
\cite{Fri,95Nel} in terms of the simultaneous structure functions of
velocity differences across a scale $R$:
\begin{equation}
\tilde S_n(R) \equiv \left<|{\bf u}({\bf r}+{\bf R})-{\bf u
}({\bf r})|^n\right>
\simeq (\bar\epsilon R)^{n/3}
\left({L \over R}\right)^{\delta_n}
\ , \label{Sn}
\end{equation}
where $\left<\dots\right>$ stands for a suitably defined ensemble
average, $\bar\epsilon$ is the mean energy flux per unit time per unit
mass, and $\delta_n$ is the deviation of the scaling exponent
$\zeta_n$ of the structure function from the Kolmogorov 1941 (K41)
prediction $\zeta_n\equiv n/3-\delta_n$. Since K41 follows from
dimensional analysis \cite{41Kol}, deviations require a
renormalization scale, and it is accepted \cite
{Fri,95Nel,87MS,93BCTBM,93BCBRCT} that in $\tilde S_n(R)$ it is the
outer scale of turbulence $L$ that serves this purpose. The same
renormalization scale appears in the correlation function of the
energy dissipation rate $\epsilon({\bf r},t)$ ( which is roughly
$\nu|\nabla {\bf u}({\bf r},t)|^2$ with $\nu$ the kinematic
viscosity)\cite{93SK}:
\begin{equation}
\tilde K_{\epsilon\epsilon}(R) = \left< (\epsilon ({\bf r}+{\bf
R})-\bar\epsilon)
(\epsilon ({\bf r}) -\bar\epsilon)\right> \simeq\bar\epsilon^2\left({L
\over R}\right)^{\mu}
\ , \label{Kee}
\end{equation}
where $\mu$ is known as the ``intermittency exponent" \cite{93SK}. The
appearance of the outer renormalization scale in these quantities has
been correctly interpreted as a failure of the K41 basic assumption of
inertial range scaling. The aim of this Letter is to discuss infinite
sets of local turbulent fields whose correlation functions require
{\em two} simultaneous renormalization scales, $L$ and $\eta$ where
$\eta$ is the inner (viscous) scale. These correlation functions
demonstrate that K41 fails doubly, once because of infrared and once
due to ultraviolet anomalies. This double anomaly results, in addition
to an infinite set of multiscaling exponents $\zeta_n$, with a second
infinite set of exponents that are denoted here as $\beta_l$.  The
phenomenon occurs in a similar fashion in scalar turbulent advection
and in Navier-Stokes turbulence. Since the development of the ideas is
simpler in the case of scalar fields we will present them in the
context of scalar advection and generalize later to turbulent vector
fields. In all cases the considerations are based on the
fusion\cite{96LP} of two or more near-by coordinates as is shown next.

First we generate local fields that originate from the fusion of two
points.  Consider for that a turbulent scalar field $T({\bf r},t)$ and
the product of two such fields at two adjacent points
\begin{equation}
\Psi(\bbox{\rho},{\bf r})\equiv T({\bf r}+\bbox{\rho}/2)
 T({\bf r}-\bbox{\rho}/2)
\ . \label{defPsi}
\end{equation}
It is advantageous to represent this  field as a multipole expansion
\begin{eqnarray}
\Psi (\bbox{\rho}, {\bf r} ) &=&
\sum_{l=0}^\infty  \Psi_l (\bbox{\rho}, {\bf r} ) \ ,
\label{multipole}\\
\Psi_l(\bbox{\rho},{\bf r} )  &=& \sum_{m=-l}^l Y_{lm} ( \hat
{\bbox{\rho}})
\int \Psi(\rho\hat{\bbox {\xi}},{\bf r})
 Y_{lm} ( \hat {\bbox {\xi}}) d \hat{\bbox {\xi}}  \  . \label{psil}
\end{eqnarray}
Here $\hat {\bbox {\rho}}=\bbox {\rho}/\rho$ and $\hat{\bbox{\xi}}$
are unit vectors. The orthonormal spherical harmonics
$Y_{lm}(\hat{\bbox {\rho}})$ are the eigenfunctions of the angular
momentum operator $\hat L=-i \bbox{\rho}\times \bbox{\nabla}$ which
depends only on the direction of $\bbox{\rho}$.:
\begin{equation}
\hat {L}^2Y_{lm}(\hat{\bbox {\rho}})=l(l+1)Y_{lm}(\hat{\bbox {\rho}} )
\ .
\label{eigen}
\end{equation}
Next we wish to represent $\Psi_l(\bbox{\rho},{\bf r})$ in terms of
(infinitely many) local fields which depend on ${\bf r}$ only. To this
aim we expand $\Psi(\bbox{\rho},{\bf r})$ in a Taylor series in
$\bbox\rho$.  This turns Eq.(\ref{psil}) to
\begin{eqnarray}
\Psi_l(\bbox{\rho},{\bf r}) &\equiv&
\sum_{m=-l}^l Y_{lm}(\hat{\bbox{\rho}})
\int d \hat{\bbox{\xi}} Y_{lm} (\hat{\bbox {\xi}} )
\nonumber\\
&\times&\sum_{n=0}^\infty 
{\rho^{2n}(\hat{\bbox{\xi}}\cdot{\bbox{\nabla}}')^{2n} \over (2n)!}
\Psi({\bbox{\rho}}',{\bf r})\Big |_{\rho'=0} \  .
 \label{psilex}
\end{eqnarray}
Here and below the operator $\nabla'_{\alpha}=\partial/ \partial
\rho'_\alpha$.  Note that we have only even $n$ orders since our field
$\Psi(\bbox{\rho},{\bf r})$ is even in $\bbox\rho$. Performing the
angular integrations we end up with
\begin{equation}
\Psi_l(\bbox{\rho},{\bf r}) =\hat A_l\sum_{n=l/2}^\infty {a_{2n,l}\over
(2n)!}
(\rho^2\nabla'^2)^{n-l/2}  \Psi(\bbox{\rho}',{\bf r})\Big|_{\rho'=0} \ .
\label{Al}
\end{equation}
The operators $\hat A_l$ in this equations are related to the
irreducible representation of the SO(3) group in a manner that will
clarify soon. For the first values of $l$ they are \FL
\begin{eqnarray}
\hat A_0 &=& (\bbox\rho\cdot\bbox{\nabla}')^0 \equiv 1\ , \  \
\hat A_2 = (\bbox\rho\cdot\bbox{\nabla}')^2-{1\over 3}\rho^2\nabla'^2\ ,
\nonumber\\
\hat A_4 &=&(\bbox\rho\cdot\bbox{\nabla}')^4-{6\over
7}\rho^2\nabla'^2(\bbox\rho\cdot\bbox{\nabla}')^2+{3\over 35}
\rho^4\nabla'^4  \ .\label{A4}
\end{eqnarray}
The coefficients are determined by requiring orthogonality of $A_l$ to
all $A_{l'}$ with $l'< l$ in the sense of integrating $A_lA_{l'}$ over
the direction of $\bbox\rho$. In fact the coefficients of the
$(\bbox{\rho}\cdot\bbox{\nabla}')^p$ term in $A_l$ are those of $x^p$
in the Legendre polynomial $P_l(x)$ \cite{AS}. On the other hand, the
coefficients $a_{2n,l}$ in (\ref{Al}) are determined by requiring that
the $\sum_l{\Psi_l}$ will agree with the original Taylor expansion
(\ref{psil}) which has only orders of $(\bbox\rho\cdot \bbox\nabla')$.
This means for example that $a_{p,p}=1$ for any $p$, $a_{2,0}=1/3$,
$a_{4,0}=-3/35$, etc.  Equation (\ref{Al}) can be rearranged in the
form
\begin{eqnarray}
&&\Psi_l(\bbox{\rho},{\bf r})
=\rho_{\alpha_1}\rho_{\alpha_2}\dots\rho_{\alpha_l}
\sum_{p=0}^\infty{a_{l+2p,2p}\over(l+2p)!}\rho^{2p}L_l^{\alpha_1\alpha_2\dots
\alpha_l}({\bf r})\ , \nonumber\\
&&L_{l,p}^{\alpha_1\alpha_2\dots \alpha_l}({\bf r})\equiv
\nabla^{2p} D_l^{\alpha_1\alpha_2\dots
\alpha_l}T({\bf r}+{\bbox{\rho}\over 2})T({\bf r}-{\bbox{\rho}\over
2})\Big|_{\rho=0} \ .
\label{deflocal}
\end{eqnarray}
Here we have introduced the tensorial local fields ${\bf L}_{l,p}$ and
the
local differential operators
${\hat{\bf D}}_l({\bf r})$ which for the first values of $l$ are
\begin{eqnarray}
&&\hat D_0 = 1 \ , \ \ \hat D^{\alpha\beta}_2
  =\nabla_\alpha\nabla_\beta-{\nabla^2\over 3}\delta_{\alpha \beta}\ ,
  \nonumber\\ &&\hat
D_4^{\alpha\beta\gamma\delta}=\nabla_\alpha\nabla_\beta
  \nabla_\gamma\nabla_\delta-{\nabla^2\over7}
(\delta_{\alpha\beta}\nabla_\gamma
  \nabla_\delta+\delta_{\alpha\gamma}\nabla_\beta\nabla_\delta
\nonumber \\ &+&
  \delta_{\alpha\delta}\nabla_\beta\nabla_\gamma+
\delta_{\beta\gamma}\nabla_\alpha
  \nabla_\delta+\delta_{\beta\delta}\nabla_\alpha\nabla_\gamma
+\delta_{\gamma\ 
    delta} \nabla_\alpha\nabla_\delta)\nonumber \\ &+&{\nabla^4\over
    35}(\delta_{\alpha\beta}\delta_{\gamma\delta}
  +\delta_{\alpha\gamma}\delta_{\beta\delta}
+\delta_{\alpha\delta}\delta_{\beta
    a\gamma}) \ . \label{L4}
\end{eqnarray}
Here $\nabla_{\alpha}=\partial/ \partial \rho_\alpha$. The readers
familiar with the representations of Lie groups recognize immediately
that our local fields ${\bf L}_{l,0}$ are nothing but the $2l+1$ rank
irreducible representations of the SO(3) group\cite{LL}.  The tensor
fields thus obtained are symmetric to any pairwise exchange of
indices.  We will propose now that these gradient fields have
$\eta$-related anomalous scaling which is governed by a set of
anomalous exponents $\beta_l$.  Autocorrelation functions of these
fields, and correlation functions of these fields together with field
differences across a scale $R$ depend also on $R/L$ with exponents
determined by the set $\zeta_n$.

Denote the correlation function of the tensorial field ${\bf L}_{l,p}$
with
$2n-2$ scalar $T$-fields as ${\bf C}_{2n,l}({\bf r},{\bf r}_3,
\dots{\bf r}_{2n})$:
\begin{equation}
{\bf C}_{2n,l,p}({\bf r},{\bf r}_3,\dots{\bf r}_{2n}) \equiv
\left\langle{\bf L}_{l,p}({\bf r})T({\bf r}_3)\dots
T({\bf r}_{2n})\right\rangle \ . \label{defC}
\end{equation}
Note that in this correlation function $\bbox\rho$ does not appear.
However it is related to the standard $2n$-point correlation function
in which two coordinates (say ${\bf r}_1$ and ${\bf r}_2$) are separated
by
a small distance $\bbox\rho$.  By definition
\begin{eqnarray}
&&{\cal F}_{2n}({\bf r}+{\bbox{\rho}\over 2},{\bf r}-{\bbox{\rho}\over
2},{\bf r}_3\dots{\bf r}_{2n})\nonumber\\
&=&\left\langle \Psi(\bbox\rho,{\bf r})T({\bf r}_3) \dots
T({\bf r}_{2n})\right\rangle \ . \label{wow}
\end{eqnarray}
To connect the functions (\ref{wow}) and (\ref{defC}) we represent
${\cal F}_{2n}$ as a multipole decomposition ${\cal
  F}_{2n}=\sum_{l=0}^\infty {\cal F}_{2n,l}$. Using (\ref{psil}) we
have
\begin{eqnarray}
&&{\cal
F}_{2n,l}({\bf r}+{\bbox{\rho}\over2},{\bf r}
-{\bbox{\rho}\over2},{\bf r}_3\dots{\bf r}_
{2n})
\nonumber\\ &=&\left\langle
\Psi_l(\bbox\rho,{\bf r})T({\bf r}_3) \dots T({\bf r}_{2n})\right\rangle
\ .
\label{wow2}
\end{eqnarray}
We are interested in the scaling properties of this quantity in the
regime in which all the separations between all the coordinates
${\bf r},{\bf r}_3,\dots{\bf r}_{2n}$ are of the order of $R$.  For
$\rho\ll R$
we can write
\begin{equation}
{\cal F}_{2n,l}({\bf r}+{\bbox{\rho}\over 2},{\bf r}-{\bbox{\rho}\over
2},
{\bf r}_3\dots{\bf r}_{2n})\sim
\left ({\rho \over R}\right)^{x_l} S_{2n}(R) \  , \label{defx}
\end{equation}
where $x_l$ is a yet unknown exponent which in general may depend also
on $n$. This exponent will be found below in a particular model and
will be shown to be $n$-independent.  For $\rho$ very small we can use
(\ref{deflocal}) and (\ref{defC}) to write
\begin{eqnarray}
&&{\cal F}_{2n,l}({\bf r}+{\bbox{\rho}\over 2},{\bf r}-{\bbox{\rho}\over
2},{\bf r}_3\dots{\bf r}_{2n})=
\rho_{\alpha_1}\rho_{\alpha_2}\dots\rho_{\alpha_l}\label{almost} \\
&&\times\sum_{p=0}^\infty{a_{l+2p,2p}\over(l+2p)!}\rho^{2p}
C_{2n,l,p}^{\alpha_1\dots\alpha_l}
({\bf r},{\bf r}_3,\dots{\bf r}_{2n})\ . \nonumber
\end{eqnarray}
Finally , in the limit $\rho\ll\eta$ we use the fact that ${\cal
  F}_{2n}$ is smooth in $\rho$ up to $\rho\sim\eta$ to evaluate the
differential operator as divisions by $\eta$:
$\rho^{2p}\nabla^{2p}\sim (\rho/\eta)^ {2p}$. Accordingly we we have
in this limit
\begin{eqnarray}
&&\lim_{\rho\to 0}{\cal F}_{2n,l}({\bf r}+{\bbox{\rho}\over
2},{\bf r}-{\bbox{\rho}\over
2},{\bf r}_3\dots{\bf r}_{2n})=\label{endgeneral}\\
&&\rho_{\alpha_1}\rho_{\alpha_2}\dots\rho_{\alpha_l}
{a_{l,0}\over
l!}C_{2n,l,0}^{\alpha_1\dots\alpha_l}({\bf r},
{\bf r}_3,\dots{\bf r}_{2n})\propto\rho
^l \ .
\nonumber
\end{eqnarray}
Next we want to explore the scaling behavior of ${\cal F}_{2n,l}$ for
values of $\rho$ in the inertial range $\eta\ll\rho\ll L$. This we
cannot do in general. We need to specialize now to a particular
dynamical model. We choose Kraichnan's model of passive advection of a
scalar field $T({\bf r},t)$ by a random velocity field whose statistics
are Gaussian, and whose correlation functions are scale invariant in
space and $\delta$-correlated in time \cite{68Kra,94Kra}. The
relevance of the results to Navier-Stokes turbulence will be discussed
later.  For a scalar diffusivity $\kappa$ the dissipation field is
$\epsilon ({\bf r}) \equiv \kappa|\nabla T|^2$ and the quantities
(\ref{Sn}) and (\ref{Kee}) are replaced by \FL
\begin{eqnarray}
S_{2n}(R) &\equiv &\left<|T({\bf R})-T(0)|^{2n}\right>
\simeq \left[S_2(R)\right]^n \left({L \over R}\right)^{\delta_n}
\ , \label{SnT}\\
K_{\epsilon\epsilon}(R) &=& \left< (\epsilon ({\bf R})-\bar\epsilon)
(\epsilon (0) -\bar\epsilon)\right> \simeq\bar\epsilon^2\left({L \over
R}\right)^{\mu}
\ . \label{KeeT}
\end{eqnarray}
In the present case the scaling exponent of $S_{2n}$ is
$\zeta_{2n}=n\zeta_2-\delta_n$.

It was shown in \cite {95FGLP} that the correlation function ${\cal
  F}_{2n}$ solves a particularly simple equation when two of its
coordinates (say ${\bf r}_1$ and ${\bf r}_2$) are much closer to one
other
than all the rest.  Explicitly, for $\rho$ small the $\bbox\rho$
dependence of this function is governed by the equation
\begin{equation}
\hat{\cal B}(\bbox{\rho}){\cal F}_{2n}({\bf r}+{\bbox{\rho}\over 2},{\bf
r}-
{\bbox{\rho}\over 2},{\bf r}_3\dots{\bf r}_{2n})
=\Phi_{2n-2}({\bf r},{\bf r}_3\dots{\bf r}_{2n})
\ . \label{fused}
\end{equation}
Here $\Phi_{2n-2}({\bf r},{\bf r}_3\dots{\bf r}_{2n})$ is a homogeneous
function with scaling exponent $\zeta_{2n}-\zeta_2$.  In 3-dimensions
the operator $\hat{\cal B}(\bbox{\rho})$ is given by
\cite{68Kra,95FGLP}
\begin{equation}
\hat{\cal B}(\bbox{\rho} )\equiv H\left[ {\partial \over \rho^2\partial
\rho}
\rho^{4-\zeta_2}   {\partial\over \partial \rho}-{(4-\zeta_2)\over
2\rho^{\zeta_2}} \hat L^2\right]  \ . \label{oper}
\end{equation}
Here $H$ is a constant.  It has been shown \cite{95FGLP,95CFKL} that
the leading scaling solution for the $\bbox\rho$ dependence of
function ${\cal F}_{2n}$ is an eigenfunction of the operator
$\hat{\cal B}(\bbox{\rho} )$ with eigenvalue $0$ and thus can be
expanded in spherical harmonics:
\begin{eqnarray}
&&{\cal F}_{2n}({\bf r}+\bbox{\rho}/2,{\bf r}
-\bbox{\rho}/2,{\bf r}_3\dots{\bf r}_{2n})
\label{expand} \\
&=&\sum_{l=0}^{\infty} \sum_{m=-l}^l
A^{(2n)}_{lm}({\bf r},{\bf r}_3\dots{\bf r}_n)\rho^{\beta_l}
Y_{l,m}(\hat{\bbox {\rho}} )
\nonumber \ ,
\end{eqnarray}
where $A^{(2n)}_{lm}({\bf r},{\bf r}_3\dots {\bf r}_n)$ is a homogeneous
function whose scaling exponent is $\zeta_{2n}-\beta_l$. To compute
the exponents $\beta_l$ for $l\ne 0$ we need to find a solution of the
homogeneous part of (\ref{fused}).  By a direct substitution of
(\ref{expand}) into the LHS of (\ref{fused}) one finds
$\beta_l(\beta_l+3-\zeta_2)= (4-\zeta_2)l(l+1)/2$. Note that the LHS
of this relation originates from the radial part of the operator
$\hat{\cal B}$, whereas the RHS results from the angular part that is
proportional to ${\hat L}^2$.  Solving the quadratic equation for
$\beta_l$ we find in 3-dimensions \cite{95FGLP,95CFKL}:
\begin{equation}
\beta_l={1 \over 2}\Big[\zeta_2-3 +\sqrt{(3-\zeta_2)^2 +
2l(l+1)(4-\zeta_2)}\Big]
\ . \label{betal}
\end{equation}
The multipole decomposition of (\ref{expand}), similarly to
(\ref{multipole}) and (\ref{psil}), leads to
\begin{eqnarray}
&&{\cal
F}_{2n,l}({\bf r}+\bbox{\rho}/2,{\bf r}-\bbox{\rho}/2,{\bf r}_3
\dots{\bf r}_{2n})
=\nonumber \\
&&\rho^{\beta_l}\sum_{m=-l}^lY_{lm}(\hat{\bbox{\rho}}
)A^{(2n)}_{lm}({\bf r},{\bf r}_3\dots{\bf r}_n)
 \ . \label{good}
\end{eqnarray}
In the situations in which all the separations between the coordinates
${\bf r},{\bf r}_3\dots{\bf r}_n$ are of the same order of magnitude
$R$, and
$R\gg \rho\gg \eta$ we can write
\begin{equation}
{\cal F}_{2n,l}({\bf r}+\bbox{\rho}/2,{\bf r}-
\bbox{\rho}/2,{\bf r}_3\dots{\bf r}_{2n})
\propto
\rho^{\beta_l}R^{\zeta_{2n}-\beta_l} \ . \label{scale}
\end{equation}
Comparing with Eq.(\ref{defx}) we identify the exponent $x_l$ as
$\beta_l$
and write the final
form:
\begin{equation}
{\cal F}_{2n,l}({\bf r}+{\bbox{\rho}\over 2},{\bf r}-{\bbox{\rho}\over
2},{\bf r}_3\dots{\bf r}_{2n}) \sim
\left({\rho\over R}\right)^{\beta_l}S_{2n}(R) \ . \label{scalefinal}
\end{equation}
At this point we want to match the solution (\ref{scalefinal}) which
is valid for $\rho \gg\eta$ with the solution (\ref{endgeneral}) which
is valid for $\rho \ll \eta$.  This can be done if the solution is
varying smoothly across $\eta$ without any non-monotonic behavior.
The rigorous proof of this property is beyond the scope of this
Letter. It can be demonstrated numerically by solving the ordinary
differential equation (\ref{fused}).  Here we simply assume this.
Equating (\ref{scale}) and (\ref{endgeneral}) for $\rho=\eta$ (up to
an unknown $R$-independent coefficient) we find
\begin{equation}
{\bf C}_{2n,l,0}({\bf r},{\bf r}_3,\dots{\bf r}_{2n})
 \sim {1\over \eta^l}\left({\eta\over
R}\right)^{\beta_l}S_{2n}(R) \ , \label{fine}
\end{equation}
where we remind the reader that $R$ stands for the order of magnitude
of all the separations between the coordinates of ${\bf C}_{2n,l,0}$.
Comparing with Eq.(\ref{SnT}) we conclude that the correlation
function of ${\bf L}_{l,0}$ with any even number of $T$ fields separated
by distances of the order of $R$ depends simultaneously on two
renromalization scales, $\eta$ and $L$, and on the two sets of
anomalous exponents $\beta_l$ and $\zeta_n$.

Until now we considered correlations of one new local field with a
number of $T$ fields.  We can also examine a cross correlation of two
(generally different) local fields. Repeating the analysis one finds
\begin{equation}
\left\langle {\bf L}_{l,0}({\bf r}+{\bf R}) 
{\bf L}_{l',0}({\bf r})\right\rangle
\sim\eta^{-l+l'}\left({\eta\over
R}\right)^{\beta_l+\beta_{l'}}S_{4}(R) \ . \label{cross}
\end{equation}
We see that in general such correlations depend on the two
renormalization scales and on two sets of exponents. It is therefore
interesting to ask why this phenomenon is absent in
$K_{\epsilon\epsilon}$ which is closely related to such correlation
functions.  We note that in terms of our local fields the correlation
(\ref{KeeT}) is given by $K_{\epsilon\epsilon}(R) =
\kappa^2\left\langle{\bf L}_{0,1}({\bf r}+{\bf R})
{\bf L}_{0,1}({\bf r})\right\rangle
$ as can be checked by substituting the definition of the local
fields. This is a very special case among the correlations of the
local fields. The reason is that for $l=0$ the differential
equation(\ref{fused}) is inhomogeneous, and it is easy to see by power
counting that $\beta_0$ is replaced by $\zeta_2$. From
Eq.(\ref{deflocal}) it follows that in this case
\cite{94Kra,95FGLP,95CFKL}
\begin{equation}
K_{\epsilon\epsilon}(R) \sim {\kappa^2\over \eta^4}
\left({\eta\over R}\right)^{2\zeta_2}S_{4}(R) \sim\bar{\epsilon}^2
\left({L\over R}\right)^{2\zeta_2-\zeta_4}
\ .\label{KeeT3}
\end{equation}
In the last step we used the fact that by definition
$\bar\epsilon=-\kappa\lim_{|{\bf r}_1-{\bf r}_2|\to
\eta}\nabla_1\nabla_2{\cal F}({\bf r}_1,{\bf r}_2)$. Since
${\cal F}({\bf r}_1,{\bf r}_2)\sim|{\bf r}_1-{\bf r}_2|^{\zeta_2}$ we
get
$\bar\epsilon\propto
\kappa\eta^{\zeta_2-2}$. This leads directly to the final step in
(\ref{KeeT3}), in which the
renormalization scale $\eta$ from the correlator of $\kappa{\bf
L}_{0,1}\sim
\kappa \nabla^2 T^2$.
The deep reason for this is that this is the rate of dissipation of the
integral of motion in the
passive scalar problem. In this sense
$K_{\epsilon\epsilon}$ is unusual, and all the generic correlations
(\ref{cross}) are simultaneously
dependent on two renormalization scales.

One can generate more local fields that will have scaling properties
which may depend on new exponents. Instead of starting with the fusion
of two points we can fuse three, four or more points. Instead of
(\ref{defPsi}) we can introduce
\begin{equation}
\Psi_3(\bbox{\rho}_1,\bbox{\rho}_2,{\bf r})\equiv
T({\bf r}+\bbox{\rho}_1)T({\bf r}+\bbox{\rho}_2)
T({\bf r}-\bbox{\rho}_1-\bbox{\rho}_2),
\end{equation}
$\Psi_4\propto T^4$ etc.  Expanding these fields in Taylor series with
respect to $\bbox{\rho}_1,\bbox{\rho}_2$ etc, we can generate new sets
of local fields that contain derivatives of three, four
etc.$T$-fields. Their correlation functions will depend on the
ultra-violet exponents which appear due to three-point, four- point
etc. coalescing clusters, and on the infrared scaling exponents of
six, eight and more-point correlation functions. Of course, the actual
values of the exponents depend on the dynamical model, but the
structure of the theory is general.  To stress this generality we make
now a few comments about the Navier-Stokes problem.  In dealing with
Navier-Stokes turbulence we need to worry from the beginning about
Galilean invariance in addition to the SO(3) symmetry group. To this
aim we will consider local fields that originate from the fusion of
gradient fields.  The simplest object is
\begin{equation}
\Psi^{\alpha\beta\gamma\delta}_2(\bbox{\rho},{\bf r}) \equiv {\partial
u_\alpha({\bf r}+\bbox{\rho}/2)
\over
\partial \rho_\beta} {\partial u_\gamma({\bf r}-\bbox{\rho}/2) \over
\partial \rho_\delta} \ . \label{tensor}
\end{equation}
From this point on we can proceed following the route sketched above
for the scalar case.  Representing this field as a multipole
decomposition with respect to the direction of $\bbox\rho$, and
considering the Taylor expansion in $\bbox\rho$, we can generate
infinitely many local fields. These fields have two ${\bf u}$ fields and
as many gradients as we want to consider, starting from two. It is
interesting to notice that in the present case we have two different
vectors, i.e. $\bbox\nabla$ and ${\bf u} $ from which we can form
untisymmetric combinations, like the vorticity
$\omega_\alpha=\epsilon_{\alpha\beta\gamma}
 \partial u_\beta / \partial r_\gamma$ (where
$\epsilon_{\alpha\beta\gamma}$
 is the fully antisymmetric tensor). Consequently we will have odd as
 well as even $l$ components in this scheme.  In addition we have
 symmetric combinations of velocity derivatives like the strain tensor
 $ s_{\alpha\beta}=[\partial u_\alpha /
 \partial r_\beta+\partial u_\beta / \partial r_\alpha]/2$. In general
the
 tensor (\ref{tensor}) has 36 independent components, serving as a
 basis for a 36-dimensional reducible representation of the O(3) group
 (SO(3)+inversion). This basis may be decomposed into a set of
 irreducible bases of lower dimensions. There are two scalar fields,
 $\omega_\alpha\omega_\alpha$ and $s^2=s_{\alpha\beta}s_{\beta\alpha}$
 each of which is a basis for one-dimensional irreducible
 representation with $l=0$.  The pseudo-vector
 $s_{\alpha\beta}\omega_\beta$ is a three-dimensional basis for an
 irreducible representation with $l=1$. There exist three traceless
 tensor fields each of which is a five-dimensional basis belonging to
 $l=2$ and taking care of $3\!\times\!5=15$ components. An example is
\begin{equation}
O_2^{\alpha\beta}({\bf r})=\omega_\alpha({\bf r})
\omega_\beta({\bf r})-\delta_{\alpha\beta}
\omega^2({\bf r})/3 \ .
\label{O2}
\end{equation}
In addition we have one 3-rank pseudo tensor corresponding to $l=3$
and one 4-rank tensor corresponding to $l=4$. The last two fields
exhaust the remaining $7+9$ components.  As in the scalar case there
are fields with all values of $l$ which are obtained when more
gradients act on our field (\ref{tensor}). Finally we can also start
with a higher number of fusing gradient fields $\partial
 u_\alpha/\partial\rho_\beta$ to generate new sets of local fields
having three, four and more velocity fields.  The exploration and
utilization of this rich structure is beyond the scope of this Letter.
It will suffice here to state that these local fields will have
correlation function with anomalous scaling properties that generally
depend on two renormalization scales and on two sets of anomalous
scaling exponents.  The correlator (\ref{Kee}) will again be special
and $\eta$-independent since it involves the rate of dissipation of
the integral of motion (energy). Correlations of fields with $l\ne 0$
will be generic.  For example the correlation of $\nu {\bf O}_2$ with
$\epsilon =2\nu s^2$ is
\begin{equation}
\nu^2\left\langle O^{\alpha\beta}_2({\bf r} +{\bf R}) s^2({\bf r}
)\right\rangle
\sim \bar\epsilon^2\left({L\over R}\right)^x
\left({\eta\over R}\right)^y \ . \label{bingo}
\end{equation}
Our guess is that $x$ is numerically close to $\mu$ and that $y$ is
numerically close to $2/3$, with an accuracy which is of the order of
the difference between $\zeta_2$ and its K41 estimate of $2/3$. We
stress however that the main point of this Letter is {\em not} the
numerical value of this or that exponent, but that normal scaling,
which is based on dimensional analysis (like K41 for Navier-Stokes
turbulence), fails doubly due to the explicit appearance of two
physically important scales, the inner {\em and} the outer
renormalization scales.

\end{document}